\documentclass[12pt]{article}
\usepackage{amssymb}
\usepackage{amsmath}
\usepackage{amsopn}
\usepackage{pifont}
\usepackage{color}
\usepackage{mathrsfs}
\usepackage{amsthm}

\numberwithin{equation}{section}

\theoremstyle{plain}
\newtheorem{thm}{Theorem}[section]
\newtheorem{prop}[thm]{Proposition}
\newtheorem{lemma}[thm]{Lemma}
\newtheorem{cor}[thm]{Corollary}
\newtheorem{Def}{Definition}[section]

\theoremstyle{definition}
\newtheorem{remark}{Remark}[section]

\def\bn{\begin{equation}}
\def\ed{\end{equation}}

\newcommand{\mage}{\color{magenta}}
\newcommand{\liashyk}[1]{{\mage$\clubsuit$}\marginpar{\framebox{{\mage \parbox{16mm}{\tiny #1}}}}}
\usepackage{cancel}

\newcommand{\so}{\scriptscriptstyle \rm I}
\newcommand{\st}{\scriptscriptstyle \rm I\hspace{-1pt}I}
\newcommand{\sth}{\scriptscriptstyle \rm I\hspace{-1pt}I\hspace{-1pt}I}

\newcommand{\bu}{\bar u}

\def\<{\langle}
\def\>{\rangle}
\newcommand{\CC}{{\mathbb C}}

\newcommand{\BB}{{\mathbb B}}

\newcommand{\ZZ}{{\mathbb Z}}

\def\r#1{(\ref{#1})}

\def\ot{\otimes}

\def\sk#1{\left(#1\right)}

\def\Pepm{{P}^\pm_e}

\def\Pfpm{{P}_f^\pm}
\def\Pfp{{P}^+_f}
\def\Pfm{{P}^-_f}

\def\Uqsl2{U_q(\widehat{\mathfrak{sl}}_2)}

\def\EE{{\rm E}}\def\FF{{\rm F}}

\def\RR{{\rm R}}
\def\LL{{\rm T}}

\def\Tt{{\rm T}}

\def\E{{\sf e}}
\def\F{{\mathcal{F}}}

\def\tLL{{\tilde{\rm T}}}

\def\qed{\hfill$\square$\medskip}

\def\bw{\bar w}
\def\proof{\noindent{\it Proof.}\ }
\def\qed{$\scriptstyle{\blacksquare}$}

\def\gle{g}
\def\gri{\tilde g}

\def\hle{h}
\def\hri{\tilde h}

\def\qed{\hfill\nobreak\hbox{$\square$}\par\medbreak}
\def\Id{\mathbb{I}}
\def\Per{\mathbb{P}}

\def\rvac{|{\rm vac}\rangle}
\def\lvac{\langle{\rm vac}|}

\def\gaf{\tilde{\mathfrak{g}}}

\def\AR{\mathbb{R}}

\def\PR{\mathbb{P}}

\def\pf{{\sf p}}

\def\glaf{\tilde{\mathfrak{gl}}_N}
\def\dfun{\gamma}
\def\FFF{\mathbb{F}}
\def\Act{\mathcal{A}}
\def\Kle{\mathsf{K}}
\def\Kri{\tilde{\mathsf{K}}}

\def\bt{\bar t}
\def\bs{\bar u}

\begin{document}

\thispagestyle{empty}
\setcounter{page}{0}

\vspace{12pt}

\begin{center}
\begin{Large}
{\bf Recurrence relations for off-shell Bethe vectors \\[2mm] 
in trigonometric integrable models}
\end{Large}

\vspace{20pt}

\begin{large}
A.~Liashyk${}^{a,b}$ and  S.~Z.~Pakuliak${}^{c,d}$
\end{large}

\vspace{10mm}

${}^a$ {\it Skolkovo Institute of Science and Technology, Moscow, Russia,}\\
${}^b$ {\it NRU Higher School of Economics, Moscow, Russia,\\
E-mail: a.liashyk@gmail.com}

\vspace{2mm}

${}^c$ {\it Bogoliubov Laboratory for Theoretical Physics, JINR\\
 Dubna, Moscow region, Russia}\\
${}^d$ {\it Landau School of Physics and Research, NRU MIPT\\
 Dolgoprudny, Moscow region, Russia\\
E-mail: stanislav.pakuliak@jinr.ru}

\end{center}


\vspace{10mm}

\noindent{\footnotesize {\bf Abstract.}
The zero modes method is applied in order to get action  of the monodromy 
matrix entries onto  off-shell Bethe vectors in quantum integrable models associated with 
$U_q(\mathfrak{gl}_N)$-invariant $\RR$-matrices. The action formulas allow
to get recurrence relations for off-shell Bethe vectors 
and for highest coefficients of the Bethe vectors scalar product.}


\section{Introduction}

Algebraic Bethe ansatz \cite{FadLH96} is a  method to describe the space of states 
of the quantum integrable models. It is applicable, in particular,  to the models defined by 
the monodromy matrices satisfying quadratic commutation relations and 
possessing special vacuum vector $\rvac$ in the space of states which is 
annihilated by the low-triangular monodromy matrix entries. 
A wide class of the quantum integrable models is related to the 
quantum groups $U_q(\mathfrak{g})$ \cite{Dr88} if structural constants of the 
quadratic commutation relations for the monodromy operators 
are $U_q(\mathfrak{g})$-invariant 
$\RR$-matrices. In this paper we consider a class of quantum trigonometric integrable models 
defined by the $U_q(\mathfrak{gl}_N)$-invariant $\RR$-matrices.
Entries of these matrices can be written as trigonometric 
functions of the parameters and this is why we call such models  {\it trigonometric}. 

Quantum affine algebras $U_q(\widehat{\mathfrak{g}})$  have several 
descriptions which use different sets of generators.
One of the description uses $\LL$-operators \cite{RS1990} such that their commutation relations 
are defined by the same $U_q(\mathfrak{g})$-invariant 
$\RR$-matrices which define monodromy matrix entries commutation relations.
This coincidence  opens a possibility to describe the space of states 
of the quantum integrable models in terms of the generators of 
the quantum affine algebras if one identifies the monodromy matrix 
of the quantum integrable model with  $\LL$-operator of $U_q(\widehat{\mathfrak{g}})$.

In \cite{HLPRS-RA} a method called {\it zero modes method} was introduced 
for the class of supersymmetric rational $\mathfrak{gl}(m|n)$-invariant 
quantum integrable models related to the super-Yangian double 
$DY(\mathfrak{gl}(m|n))$. This method uses the commutation relations 
between zero modes of the monodromy matrix entries and entries themselves 
in order to obtain the action of these entries onto off-shell Bethe vectors 
in the corresponding quantum integrable model. In the present paper 
we develop the zero modes method to the trigonometric quantum integrable 
models related to the quantum affine algebra $U_q(\widehat{\mathfrak{gl}}_N)$.

The action formulas of monodromy matrix entries onto off-shell Bethe vectors 
are more fundamental than explicit expressions for these vectors 
in terms of monodromy matrix elements. 
They can be used to investigate the physical quantities in the quantum integrable 
models such as scalar products and form-factors of the local operators 
without using explicit formulas for the Bethe vectors. 
On the other hand, 
the action formulas lead 
to the recurrent relations for the off-shell Bethe vectors, which can be 
solved to obtain explicit expressions for them.
For the off-shell Bethe 
vectors in the quantum integrable models defined by 
$U_q(\mathfrak{gl}_N)$-invariant $\RR$-matrices two particular recurrence 
relations were obtained in \cite{HLPRS18} using method of 
the hierarchical Bethe ansatz. From viewpoint of this method these two types of the 
recurrence relations are related to two different ways of embedding 
$U_q(\mathfrak{gl}_{N-1})$ monodromy matrix into $U_q(\mathfrak{gl}_{N})$ monodromy:
either in the upper-left corner or in the down-right corner. 

To describe off-shell Bethe vectors in terms of generators of the 
quantum affine algebra one has to explore the Gauss decomposition 
of the $\LL$-operators. There are different Gauss decompositions
naturally related to the embeddings mentioned above. It was shown in  
 \cite{OPS} that each of these embeddings leads  to the one 
 type recurrence relation and it is  not easy combinatorial problem to prove that 
 different type recurrence relations lead to different but equivalent 
 presentations of the same off-shell Bethe vectors. For the rational quantum integrable models 
 defined by the $\mathfrak{gl}(m|n)$-invariant $\RR$-matrices
 this problem was solved in \cite{HLPRS17}. 
 
Zero modes method developed in the present paper allows to find 
the action formulas of monodromy matrix entries onto off-shell Bethe vectors
in trigonometric integrable models. This action can be used to get 
various recurrence relations including those obtained in \cite{HLPRS18}. 
Similar results were obtained  in \cite{LP21} for the case 
of the rational $\mathfrak{o}_{2n+1}$-invariant integrable models. 

The paper is composed as follows. In section~\ref{defnot} the quantum 
loop algebra $U_q(\glaf)$ in terms of fundamental $\LL$-operators  is defined. 
Section~\ref{mres} contains two main 
results of this paper, which include single and multiple actions of monodromy 
matrix entries onto off-shell Bethe vectors and recurrence relations for them. 
Section~\ref{zmmeth} is devoted to formulation of the zero modes method, 
which is used to obtain action formulas. 
Third result of the paper -- the recurrence relations for the highest coefficients of the 
Bethe vectors scalar product is presented in the section~\ref{sec:BV}. 
 Proofs of the main propositions are 
gathered in two appendices.

\section{Definitions and notations}
\label{defnot}

In this paper we will explore
 the Cartan-Weyl generators of the quantum loop algebra $U_q(\glaf)$. 
This algebra is related to the quantum affine algebra 
by setting the central element in $U_q(\widehat{\mathfrak{gl}}_N)$ equal to zero.

\subsection{$\RR$-matrix for $U_q(\tilde{\mathfrak{gl}}_N)$}
\label{RRsect}

Let $N$ be dimension of the fundamental vector representation of the algebra 
$U_q(\mathfrak{gl}_N)$ in $\CC^N$. 
Let $\E_{ij}$ be a $N\times N$ matrix unit $(\E_{ij})_{k,l}=\delta_{ik}\delta_{jl}$ 
for $1\leq i,j,k,l\leq N$.
 We introduce functions 
\begin{equation}\label{rat-fun}
f(u,v)=\frac{qu -q^{-1}v}{u-v}\,,\quad 
\gle(u,v)=\frac{(q-q^{-1})u}{u-v}\,,\quad
\gri(u,v)=\frac{(q-q^{-1})v}{u-v}
\end{equation}
of the arbitrary complex spectral parameters   $u$ and $v$.

Define matrix $\PR(u,v)$  acting in the tensor product 
$\CC^N\ot\CC^N$ by the equality
\begin{equation}\label{PPuv}
\PR(u,v)= \sum_{1\leq i,j\leq N} \pf_{ij}(u,v)\ \E_{ij}\ot \E_{ji}\,,
\end{equation}
where 
rational functions $\pf_{ij}(u,v)$   are defined as follows
\begin{equation*}\label{p-fun}
\pf_{ij}(u,v)=\begin{cases}
f(u,v)-1,\quad &i=j\,,\\
\gle(u,v),\quad &i<j\,,\\
\gri(u,v),\quad &i>j\,.
\end{cases}
\end{equation*} 

\begin{Def}
Quantum trigonometric $U_q(\mathfrak{gl}_N)$-invariant 
$\RR$-matrix acting in the tensor product of 
two fundamental vector representations of $U_q(\mathfrak{gl}_N)$  is 
\begin{equation}\label{R-matA}
\AR(u,v)=\Id\ot\Id+\PR(u,v)\,,
\end{equation}
where $\Id=\sum_{i=1}^N\E_{ii}$ is identity matrix in $\CC^N$. 
\end{Def}

\subsection{$\RR$-matrix formulation of the algebra $U_q(\glaf)$}
\label{Aldef}

The associative algebra $U_q(\glaf)$ with unit $\mathbf{1}$ over $\CC(q)$ 
is generated by the elements $\LL^\pm_{i,j}[\pm m]$,
$1\leq i,j\leq N$, $m\in\ZZ_+$ such that 
\begin{equation}\label{restr}
\LL^+_{j,i}[0]=\LL^-_{i,j}[0]=0, \quad i<j,\quad 
\LL^+_{i,i}[0]\LL^-_{i,i}[0]=\LL^-_{i,i}[0]\LL^+_{i,i}[0]=\mathbf{1}.
\end{equation}

The generators of the algebra $U_q(\glaf)$ may be gathered into formal series 
\begin{equation}\label{series}
\LL^\pm_{i,j}(u)=\sum_{m=0}^\infty \LL^\pm_{i,j}[\pm m] u^{\mp m}
\end{equation}
and combined in the matrices 
\begin{equation}\label{Lop}
\LL^\pm(u)=\sum_{i,j=1}^N \E_{ij}\ot \LL^\pm_{i,j}(u)\in {\rm End}(\CC^N)\ot 
U_q(\glaf)[[u^{\mp 1}]]
\end{equation}
which we call  $\LL$-operators.

The commutation relations in the algebra $U_q(\glaf)$ are given 
by the standard RLL commutation relations  
\begin{equation}\label{RLL}
\RR(u,v) \cdot (\LL^\mu(u) \ot \Id)\cdot (\Id\ot \LL^\nu(v))=
(\Id\ot \LL^\nu(v))\cdot (\LL^\mu(u) \ot \Id)\cdot \RR(u,v)\,,
\end{equation}
where $\mu,\nu=\pm$  and rational functions entering $\RR$-matrix 
$\RR(u,v)$ \r{R-matA} 
 should be understood as series over $v/u$ for $\mu=+$, $\nu=-$ and 
as series over $u/v$ for $\mu=-$, $\nu=+$. For $\mu=\nu$  
these rational series can be either series over the ratio $v/u$ or the ratio  $u/v$.

The commutation relations in the algebra $U_q(\gaf)$ may be written in terms
of matrix entries \r{series}. Using explicit expression 
\r{PPuv} one gets 
\begin{equation}\label{TM-1}
[\LL^\mu_{i,j}(u),\LL^\nu_{k,l}(v)]=
\pf_{lj}(u,v)\ \LL^\nu_{k,j}(v)\LL^\mu_{i,l}(u)-\pf_{ik}(u,v)\ \LL^\mu_{k,j}(u)\LL^\nu_{i,l}(v)\,.
\end{equation} 
It follows from the commutation relations \r{RLL} or \r{TM-1} that modes $\LL^{\pm}_{i,j}[m]$,  $m\geq 0$ form Borel subalgebras $U^\pm_{q}(\glaf)\subset U_q(\glaf)$.

\begin{remark}\label{rem1}
One can  check that the restrictions to the 
zero mode generators \r{restr} are consistent  with the commutation relations 
\r{TM-1}. 
\end{remark}

\subsection{Gauss coordinates and the currents} 
\label{GC}

Let $\mathcal{H}$ be a representation space of the algebra $U_q(\glaf)$ which 
possesses a vector $\rvac$ with following properties
\begin{equation}\label{vac-def}
\LL^\pm_{i,j}(u)\rvac=0,\quad i>j,\qquad \LL^\pm_{i,i}(u)\rvac=\lambda^\pm_i(u)\rvac\,.
\end{equation}

In what follows we will use Gauss decomposition of the  $\LL$-operators 
of the algebra $U_q(\glaf)$ \cite{DF93}
\begin{equation}\label{Gauss1}
\LL^\pm_{i,j}(u)=\sum_{\ell\leq{\rm min}(i,j)} \FF^\pm_{j,\ell}(u)\ k^\pm_\ell(u)\ \EE^\pm_{\ell,i}(u)\,.
\end{equation}
We call series $\FF^\pm_{j,i}(u)$, $\EE^\pm_{i,j}(u)$ for $1\leq i< j\leq N$ 
and $k^\pm_{\ell}(u)$ for $1\leq\ell\leq N$
the {\it Gauss coordinates}.
Properties of the vacuum vector \r{vac-def} are 
 translated  to 
\begin{equation*}\label{vac-def-GC}
\EE^\pm_{i,j}(u)\rvac=0,\quad i<j,\qquad k^\pm_j(u)\rvac=\lambda^\pm_j(u)\rvac\,.
\end{equation*}
In  \r{Gauss1} 
we assume that  $\FF^\pm_{i,i}(u)=\EE^\pm_{i,i}(u)=1$ for 
$1\leq i\leq N$. Series expansion of the  $\LL$-operators \r{series} 
imply following expansions of the Gauss coordinates
\begin{equation}\label{GCser}
\FF^\pm_{j,i}(u)=\sum_{m=0}^\infty \FF^\pm_{j,i}[\pm m]u^{\mp m},\quad
\EE^\pm_{i,j}(u)=\sum_{m=0}^\infty \EE^\pm_{i,j}[\pm m]u^{\mp m},\quad
k^\pm_{\ell}(u)=\sum_{m=0}^\infty k^\pm_{\ell}[\pm m]u^{\mp m}.
\end{equation}
According to \r{restr} and Gauss decomposition \r{Gauss1} $\FF^-_{j,i}[0]=0$ 
and $\EE^+_{i,j}[0]=0$.

The commutation relations between Gauss coordinates imply the commutation 
relations in the algebra $U_q(\glaf)$ in terms of the currents \cite{DF93,D88}
\begin{equation}\label{glNcurr} 
\begin{split}
F_i(u)&=\FF^+_{i+1,i}(u)-\FF^-_{i+1,i}(u)=\sum_{\ell\in\ZZ}{{\rm sign}(\ell)}\FF_{i+1,i}[\ell]u^{-\ell} ,\\
E_i(u)&=\EE^+_{i,i+1}(u)-\EE^-_{i,i+1}(u)=-\sum_{\ell\in\ZZ}{{\rm sign}(-\ell)}\EE_{i,i+1}[\ell]u^{-\ell},
\end{split}
\end{equation}
where sign function ${\rm sign}(\ell)$ is defined as 
\begin{equation}\label{sign}
{\rm sign}(\ell)=\begin{cases}+1,\quad&\ell\geq 0\,,\\
-1,\quad&\ell<0\,.
\end{cases}
\end{equation}

The nontrivial commutation relations between currents are 
\begin{equation*}\label{kiFA}
\begin{split}
k^{\pm}_i(u) F_i(v) k^{\pm}_i(u)^{-1}&= \frac{q^{-1}u -qv}{u-v}\   F_i(v),\\
k^{\pm}_{i+1}(u)F_i(v)k^{\pm}_{i+1}(u)^{-1}&= \frac{qu -q^{-1}v}{u-v}\   F_i(v),
\end{split}
\end{equation*}
\begin{equation}\label{kEFA}
\begin{split}
k^{\pm}_i(u)^{-1}E_i(v)k^{\pm}_i(u)&=\frac{q^{-1}u -qv}{u-v}\   E_i(v),\\
k^{\pm}_{i+1}(u)^{-1}E_i(v)k^{\pm}_{i+1}(u)&=\frac{qu -q^{-1}v}{u-v}\  E_i(v),
\end{split}
\end{equation}
\begin{equation*}\label{FiFiA}
(q^{-1}u -qv)\ F_i(u)F_i(v)=  (qu -q^{-1}v)\  F_i(v)F_i(u),
\end{equation*}
\begin{equation*}\label{EiEiA}
(qu -q^{-1}v)\ E_i(u) E_i(v)=  (q^{-1}u -qv)\  E_i(v) E_i(u),
\end{equation*}
\begin{equation*}\label{FiFiiA}
(u-v)\ F_i(u)F_{i+1}(v)= (q^{-1}u-qv)\ F_{i+1}(v)F_i(u),
\end{equation*}
\begin{equation*}\label{EiEiiA}
(q^{-1}u-qv)\ E_i(u)E_{i+1}(v)= (u-v)\  E_{i+1}(v)E_i(u),
\end{equation*}
\begin{equation}\label{EFA}
[E_i(u),F_j(v)]=\delta_{i,j}\ (q-q^{-1}) 
\delta(u,v)\Big(k^-_{i+1}(v)\,k^-_{i}(v)^{-1}-k^+_{i+1}(u)\,k^+_{i}(u)^{-1}\Big)
\end{equation}
and Serre  relations for the currents $E_i(u)$ and $F_i(u)$ \cite{DF93,D88}.

In \r{EFA} the multiplicative delta function is defined by the formal series 
\begin{equation*}\label{delta}
\delta(u,v)=\sum_{\ell\in\ZZ} \frac{u^\ell}{v^\ell}
\end{equation*}
which satisfy the property
\begin{equation*}\label{delta-prop}
\delta(u,v)G(u)=\delta(u,v)G(v)
\end{equation*}
for any formal series $G(u)$.

\begin{remark}\label{for-ser}
The equalities \r{kEFA} should be understood in a sense of equalities 
between formal series. It means that these commutation relations should be understood
as infinite set of equalities between modes of the currents which appear after equating 
the coefficients at all powers $u^\ell v^{\ell'}$ for $\ell,\ell'\in\ZZ$. 
The rational functions in the commutation relations  \r{kEFA}
should be understood as series over powers of $v/u$ in the relations containing 
the current $k^+_j(u)$ and over powers of $u/v$ in the relations with 
the current $k^-_j(u)$.
\end{remark}

\subsection{Sets of parameters and their partitions}

It is known \cite{KhT93} that the Cartan-Weyl generators of $U_q(\glaf)$ 
can be identified with the generators of the currents $F_i(u)$, $E_i(u)$ and $k^\pm_\ell(u)$.
Off-shell Bethe vectors which will be defined in the next section through 
these current generators  \cite{KhP-Kyoto} depend on the 
sets of the Bethe parameters.  One needs $N-1$ types of the parameters $t^\ell_a$
for $\ell=1,\ldots,N-1$ and $a=1,\ldots,r_\ell$
to describe off-shell Bethe vectors for the quantum integrable model 
associated with $U_q(\mathfrak{gl}_N)$-invariant $\RR$-matrix. 
Superscript $\ell$ in $t^\ell_a$ denotes the type of the parameter while 
subscript $a$ counts the number of the parameters of the same type. 

We will collect parameters of the same type in the sets $\bar t^\ell=\{t^\ell_1,\ldots,t^\ell_{r_\ell}\}$
with cardinalities $|\bar t^\ell|=r_\ell$ and 
will denote collection of these sets as $\bar t=\{\bar t^{1},\ldots,\bar t^{N-1}\}$. 
If non-negative number $r_\ell=0$ vanishes for some $\ell$ it means that the corresponding set 
$\bar t^\ell=\varnothing$ is empty. The set $\bar t^\ell_a$ is by definition  the set 
$\bar t^\ell\setminus\{t^\ell_a\}$ of the cardinality $|\bar t^\ell_a|=r_\ell-1$.

Our results are formulated as sums over partitions of the sets $\bar t^\ell$ into 
several nonintersecting subsets. We denote these partitions as 
$\{\bar t^\ell_{\so},\bar t^\ell_{\st}\}\vdash \bar t^\ell$ such that 
$|\bar t^\ell_{\so}|+|\bar t^\ell_{\st}|=|\bar t^\ell|$.

For any scalar  functions or commuting operators 
of one or two variables we will use  notation that 
$\lambda_j(\bar u)$ or $f(\bar u,\bar v)$ means the product of the corresponding 
functions over the elements of the sets $\bar u$ and $\bar v$:
\begin{equation*}
\lambda^+_j(\bar u)=\prod_{s=1}^{|\bar u|}\lambda^+_j(u_s),\quad
f(\bar u,\bar v)=\prod_{s=1}^{|\bar u|}\prod_{p=1}^{|\bar v|}f(u_s,v_p),\quad
\LL_{i,j}^+(\bar z)=\prod_{s=1}^{|\bar z|}\LL_{i,j}^+(z_s).
\end{equation*}
If any of these sets is empty then 
the corresponding product is equal to 1 by definition. For example, $f(\bar u,\varnothing)\equiv 1$.

\section{Bethe vectors and main results}
\label{mres}

Monodromy matrix $\Tt(z)$ of 
any quantum integrable model with $U_q(\mathfrak{gl}_N)$-invariant $\RR$-matrix 
satisfy the commutation relation \r{RLL} or \r{TM-1} for $\mu=\nu$. We identify 
$\Tt(z)\equiv\LL^+(z)$. $\RR\LL\LL$ commutation relations \r{RLL} imply that transfer matrix 
\begin{equation}\label{trans}
\mathfrak{t}(z)=\sum_{i=1}^N \Tt_{i,i}(z)
\end{equation}
commutes for different values of the spectral parameters. The goal 
of the algebraic Bethe ansatz is to describe the solution of the eigenvalue 
problem 
\begin{equation}\label{e-pro}
\mathfrak{t}(z)\cdot \BB(\bar t)=\tau(z;\bar t)\ \BB(\bar t)\,,
\end{equation} 
where 
\begin{equation}\label{e-val}
\tau(z;\bar t)=\sum_{i=1}^N\lambda_i(z) f(z,\bar t^{i-1})f(\bar t^i,z)
\end{equation}
and 
$\bar t$ is a set of Bethe parameters $\{\bar t^1,\ldots,\bar t^{N-1}\}$.
The boundary sets $\bar t^{0} = \bar t^{N}= \varnothing$ appearing in \r{e-val} are empty.
The functions $\lambda_i(z)\equiv \lambda^+_i(z)$
are free functional parameters. In each concrete integrable model these 
parameters are fixed to some functions. 
 Let $\beta_i(z)$ be a ratio of neighboring 
functional parameters 
\begin{equation}\label{beta}
\beta_i(z)=\frac{\lambda_{i+1}(z)}{\lambda_i(z)},\quad i=1,\ldots,N-1\,.
\end{equation}
The parameters of the Bethe vectors should satisfy  Bethe equations
\begin{equation}\label{BE}
\beta_i(t^i_\ell)=
\frac{\lambda_{i+1}(t^i_\ell)}{\lambda_{i}(t^i_\ell)}=
\frac{f(\bar t^i_\ell,t^i_\ell)}{f(t^i_\ell,\bar t^i_\ell)}\ \frac{f(t^i_\ell,\bar t^{i-1})}{f(\bar t^{i+1},t^i_\ell)}
\end{equation}
in order to fulfill eigenvalue problem \r{e-pro}. Such Bethe vectors 
are called {\it on-shell}. 

If Bethe parameters are free then Bethe vectors are called {\it off-shell} and have the structure 
described by hierarchical Bethe ansatz. Off-shell Bethe vectors can be described in terms 
of the Cartan-Weyl generators of the quantum loop  algebra $U_q(\glaf)$.
In this case $\BB(\bar t)\in\mathcal{H}$ are vectors in the representation space $\mathcal{H}$
of the algebra $U_q(\glaf)$. 

To describe this relation one has to consider  Borel subalgebras of 
$U^\pm_q(\glaf)\subset U_q(\glaf)$
formed  by the modes of $\LL$-operators $\LL^\pm(u)$ and 
an alternative Borel subalgebras $U_F$ and $U_E$ formed by the 
modes of the currents $F_i(u)$, $k^+_j(u)$ and $E_i(u)$, $k^-_j(u)$ respectively. 
One can define projections $\Pfpm$ and $\Pepm$ onto intersections 
$U^\pm_F=U_F\cap U^\pm_q(\glaf)$ and $U^\pm_E=U_E\cap U^\pm_q(\glaf)$. 
These projections act on the Borel subalgebras  $U_F$ and $U_E$.

Detailed investigation of these projections for  
quantum affine algebra  $U_q(\widehat{\mathfrak{gl}}_N)$ 
was performed in \cite{EKhP07}.
For the purpose of the present paper we may understood projections  $\Pfpm$
acting onto product of the simple roots currents $F_i(t)$ as follows. To calculate 
projection from the product of these currents one has substitute each 
current by the difference of the Gauss coordinates \r{glNcurr} and then use 
commutation relations between them to 'normal' order all monomials such that 
all 'negative' Gauss coordinates $\FF_{j,i}^-(t)$ are on the left of 
all 'positive' coordinates $\FF_{j,i}^+(t')$. 
Although original expressions is a collection of monomials composed from 
the  Gauss coordinates $\FF^\pm_{i+1,i}(t)$ only, the higher Gauss 
coordinates $\FF^\pm_{j,i}(t)$ for $j>i+1$ will appear due to this normal ordering 
process.
Then application of the projection 
$\Pfp$ means removing of all monomials which have at least one 'negative' 
coordinate on the left. Analogously, application of the projection $\Pfm$
means removing of all monomials which have at least one 'positive' 
coordinate on the right. 
Projections $\Pepm$ acting onto product of the currents $E_i(t)$ 
can be understood similarly according to the cyclic ordering of the 
Cartan-Weyl generators (see details in \cite{EKhP07,LP21a}).
In \cite{KhP-Kyoto} more effective methods 
to calculate such projections were developed. We address an interested 
reader to this paper and reference therein.

Let us introduce the ordered products of the simple root currents 
$\F_i(\bar t^i)$ 
\begin{equation}\label{orcp}
\F_i(\bar t^i)= \prod_{\ell<\ell'}^{r_i}f( t^i_{\ell'},t^i_{\ell}) F_i(t^i_1)F_i(t^i_1)\cdots F_i(t^i_{r_i})\,.
\end{equation}
Each $\F_i(\bar t^i)$ 
is obviously symmetric with respect to permutations of the elements in the set $\bar t^i$
due to the commutation relations \r{kEFA}.
In order to express off-shell Bethe vectors in terms of the Cartan-Weyl generators 
of the algebra $U_q(\glaf)$ we define the  normalized ordered product  
\begin{equation}\label{sopc}
\FFF(\bar t)=\prod_{i=1}^{N-2}f(\bar t^{i+1},\bar t^{i})^{-1}\ 
\F_1(\bar t^1)\F_2(\bar t^2)\cdots \F_{N-1}(\bar t^{N-1})
\end{equation}
We call projection of this product of currents $\Pfp\Big(\FFF(\bar t)\Big)$
off-shell pre-Bethe vector and 
 off-shell Bethe vector $\BB(\bar t)$ itself is \cite {KhP-Kyoto}
\begin{equation}\label{BV}
\BB(\bar t)=\BB(\bar t^1,\bar t^2,\ldots,\bar t^{N-1})=\Pfp\Big(\FFF(\bar t)\Big)\rvac\,.
\end{equation}

The commutation relations of the currents and properties 
of the projections imply that if any of the set 
of the Bethe parameters $\bar t^i=\varnothing$ is empty  then pre-Bethe vector 
factorizes into product of pre-Bethe vectors for  $U_q(\tilde{\mathfrak{gl}}_{i})$
and $U_q(\tilde{\mathfrak{gl}}_{N-i})$ algebras
\begin{equation*}
\Pfp\Big(\FFF(\bar t^1,\ldots,\bar t^{i-1},\varnothing,\bar t^{i+1},\ldots,\bar t^{N-1})\Big)=
\Pfp\Big(\FFF(\bar t^1,\ldots,\bar t^{i-1})\Big)\cdot 
\Pfp\Big(\FFF(\bar t^{i+1},\ldots,\bar t^{N-1})\Big)\,.
\end{equation*}
In particular, when $i=1$ or $i=N-1$ the off-shell Bethe vectors 
$\BB(\varnothing,\bar t^{2},\ldots,\bar t^{N-1})$ and 
$\BB(\bar t^1,\ldots,\bar t^{N-2},\varnothing)$ are 
 $U_q(\tilde{\mathfrak{gl}}_{N-1})$ Bethe vectors.

\subsection{Action formulas}

In addition to the rational functions \r{rat-fun} we introduce the functions
\begin{equation}\label{h-func}
\hle(u,v)=\frac{f(u,v)}{\gle(u,v)},\quad \hri(u,v)=\frac{f(u,v)}{\gri(u,v)}
\end{equation}
which satisfy the properties 
\begin{equation}\label{h-p1}
1-\frac{q^{-1}}{f(u,v)}=\frac{1}{\hle(u,v)}\quad\mbox{or}\quad \hle(u,v)-\frac{q^{-1}}{\gle(u,v)}=1
\end{equation}
and 
\begin{equation}\label{h-p2}
1-\frac{q}{f(u,v)}=\frac{1}{\hri(u,v)}\quad
\mbox{or}\quad \hri(u,v)-\frac{q}{\gri(u,v)}=1\,.
\end{equation}

For two sets $\bar x$ and $\bar y$ of the same cardinalities $|\bar x|=|\bar y|=n$
one can define the {\it  Izergin determinant} 
\begin{equation}\label{Ize}
\Kle(\bar x|\bar y)=
\frac{\prod_{i=1}^n x_i\prod_{1\leq i,j\leq n}(qx_i-q^{-1}y_j)}{\prod_{1\leq i<j\leq n}(x_i-x_j)(y_j-y_i)}
\ \det\left[\frac{q-q^{-1}}{(x_i-y_j)(qx_i-q^{-1}y_j)} \right]
\end{equation}
and 
\begin{equation}\label{Izri}
\Kri(\bar x|\bar y)=\prod_{i=1}^n \frac{y_i}{x_i}\ \Kle(\bar x|\bar y)\,.
\end{equation}
Note that for $n=1$
\begin{equation*}
\Kle(x|y)=\gle(x,y)\quad\mbox{and}\quad \Kri(x|y)=\gri(x,y)\,.
\end{equation*}

First main result of the  paper can be formulated as following
\begin{prop}\label{mulac}
Let $\bar z=\{z_1,\ldots,z_r\}$ be a set of arbitrary parameters of cardinality $|\bar z|=r$.
Then, the multiple action of monodromy matrix elements $\Tt_{i,j}(\bar z)$ 
 onto off-shell Bethe vector $\BB(\bar t)$ is given by the formula 
\begin{equation}\label{mac2}
\begin{split}
\Tt_{i,j}(\bar z) \mathbb{B}(\bar t) &=  \lambda_{1}(\bar z)  
      \sum_{{\rm part}} \mathbb{B}(\bar w_{\st} ) 
      \frac{\prod_{p=j}^{i-1} f(\bar w^p_{\so},\bar w^p_{\sth})}
       {\prod_{p=j}^{i-2} f(\bar w^{p+1}_{\so},\bar w^p_{\sth}) } \times  \\
&\times    \prod_{p=1}^{i-1} \frac{\beta_p(w^p_{\so})
\Kle(\bar w^p_{\so}|\bar w^{p-1}_{\so}) f(\bar w^p_{\so},\bar w^p_{\st})}
   {f(\bar w^p_{\so}, \bar w^{p-1}_{\so}) f(\bar w^p_{\so},\bar w^{p-1}_{\st})}
	\prod_{p=j}^{N-1}      
	\frac{\Kri(\bar w^{p+1}_{\sth}| \bar w^{p}_{\sth})  f(\bar w^p_{\st},\bar w^p_{\sth})   }
	{f(\bar w^{p+1}_{\sth}, \bar w^{p}_{\sth})f(\bar w^{p+1}_{\st}, \bar w^{p}_{\sth})}        
 \end{split}      
    \end{equation}
 where 
sum in \r{ac2} goes over partitions described below:
\begin{itemize}
\item     $\bar w^p = \{\bar z, \bar t^p\}$.
 These sets will be divided in subsets $\{\bar w_{\so}^p,\bar w_{\st}^p, \bar w_{\sth}^p\}
 \vdash \bar w^p$.
\item   Boundary conditions: $\bar w^0_{\so} = \bar w^{N}_{\sth} = 
\{\bar z\}$, $\bar w^0_{\st}=\bar w^0_{\sth} = \bar w^{N}_{\so}=\bar w^{N}_{\st} =  \varnothing$.
\item  Subsets $\bar w^p_{\so}$ are non empty only for $p<i$ and have cardinality 
$|\bw^s_{\so}| = r$.
\item Subsets $\bar w^p_{\sth}$ are non empty only for $p\ge j$ 
and have cardinality $|\bw^s_{\sth}| = r$.
\end{itemize}
\end{prop}

This proposition is a direct consequence of the single action formulated in the following 
\begin{prop}\label{sinac}
The single monodromy matrix entry $\Tt_{i,j}(z)$ action onto off-shell Bethe vector 
is given by the  exporession
\begin{equation}\label{ac2}
\Tt_{i,j}(z) \mathbb{B}(\bar t) =  \lambda_{1}(z)  
      \sum_{{\rm part}} \mathbb{B}(\bar w_{\st} ) \Act_{i,j}(\bar w_{\so};\bar w_{\st};\bar w_{\sth})\,,
\end{equation}
where
\begin{equation}\label{ac22}
\begin{split}
      \Act_{i,j}(\bar w_{\so};\bar w_{\st};\bar w_{\sth})  &=
      \frac{\prod_{p=j}^{i-1} f(\bar w^p_{\so},\bar w^p_{\sth})}
       {\prod_{p=j}^{i-2} f(\bar w^{p+1}_{\so},\bar w^p_{\sth}) } \times  \\
&\times    \prod_{p=1}^{i-1} \frac{\beta_p(\bar w^p_{\so}) f(\bar w^p_{\so},\bar w^p_{\st})}
   {\hle(\bar w^p_{\so}, \bar w^{p-1}_{\so}) f(\bar w^p_{\so},\bar w^{p-1}_{\st})}
	\prod_{p=j}^{N-1}      \frac{  f(\bar w^p_{\st},\bar w^p_{\sth})   }
	{\hri(\bar w^{p+1}_{\sth}, \bar w^{p}_{\sth})f(\bar w^{p+1}_{\st}, \bar w^{p}_{\sth})}        
 \end{split}      
    \end{equation}
 and
sum in \r{ac2} goes over partitions described in proposition~\ref{mulac} for $r=1$.
\end{prop}

Proofs of the propositions~\ref{sinac}  can be found in the appendices~\ref{ApA}.
Proof of proposition~\ref{mulac} is similar to the proof of 
analogous statement given in appendix~B
of the paper \cite{HLPRS-RA} and based on the properties of the Izergin 
determinant \r{Ize}. \qed

\begin{cor}
For $i\leq j$ the action of the matrix entries $\Tt_{i,j}(z)$ onto off-shell Bethe vectors
is simplified to 
\begin{equation}\label{acsim}
\Tt_{i,j}(z) \mathbb{B}(\bar t) =  \lambda_{1}(z)  
      \sum_{{\rm part}} \mathbb{B}(\bar w_{\st} )
       \prod_{p=1}^{i-1} \frac{\beta_p(\bar w^p_{\so}) f(\bar w^p_{\so},\bar w^p_{\st})}
   {\hle(\bar w^p_{\so}, \bar w^{p-1}_{\so}) f(\bar w^p_{\so},\bar w^{p-1}_{\st})}
	\prod_{p=j}^{N-1}      \frac{  f(\bar w^p_{\st},\bar w^p_{\sth})   }
	{\hri(\bar w^{p+1}_{\sth}, \bar w^{p}_{\sth})f(\bar w^{p+1}_{\st}, \bar w^{p}_{\sth})}  . 
\end{equation}
\end{cor}

\subsection{Recurrence relations}

To formulate the second main result of this paper  
we introduce the notation $\{\bar t^s\}_i^j$ which  
means collection of the sets $\{\bar t^i,\bar t^{i+1},\ldots,\bar t^j\}$.  
\begin{prop}\label{rr1var}
Off-shell Bethe vectors $\BB(\bar t)$ satisfy the recurrent relations
\begin{equation}\label{c7}
\begin{split}
&\BB(\{\bar t^s\}_1^{\ell-1},\{z,\bar t^\ell\},\{\bar t^s\}_{\ell+1}^{N-1})=\\
&\quad=\sum_{\rm part}
\sum_{i=1}^\ell\sum_{j=l+1}^{N}\frac{\Tt_{i,j}(z)}{\lambda_\ell(z)}
\frac{\BB(\{\bar t^s\}_1^{i-1},\{\bar t^s_{\st}\}_i^{\ell-1},
\bar t^\ell,\{\bar t^s_{\st}\}_{\ell+1}^{j-1},\{\bar t^s\}_{j}^{N-1}))}
{f(z,\bar t^{\ell-1})f(\bar t^{\ell+1},z)}\times\\
&\qquad\times \prod_{p=i}^{\ell-1} \frac{\beta_p(\bar t^p_{\so})
\gri(\bar t^{p+1}_{\so},\bar t^{p}_{\so})f(\bar t^p_{\so},\bar t^p_{\st})}
{f(\bar t^{p}_{\so},\bar t^{p-1})}\ \prod_{p=\ell+1}^{j-1}
\frac{\gle(\bar t^p_{\sth},\bar t^{p-1}_{\sth})f(\bar t^p_{\st},\bar t^p_{\sth})}
{ f(\bar t^{p+1},\bar t^{p}_{\sth})}\,,
\end{split}
\end{equation}
where sum goes over partitions $\{\bar t^p_{\so},\bar t^p_{\st}\}\vdash \bar t^p$
for $p=i,\ldots,\ell-1$ and partitions $\{\bar t^p_{\st},\bar t^p_{\sth}\}\vdash \bar t^p$
for $p=\ell+1,\ldots,j-1$ such that $|\bar t^p_{\so}|=1$ 
for $p=i,\ldots,\ell$ and $|\bar t^p_{\sth}|=1$ 
for $p=\ell,\ldots,j-1$ with fixed boundary partitions $\bar t^\ell_{\so}=\bar t^\ell_{\sth}=\{z\}$
and $\bar t^0=\bar t^N=\varnothing$.
\end{prop} 

Proof of the proposition~\ref{rr1var} can be found in the appendix~\ref{ApC}.

Recurrence relation \r{c7} is written in assumption that all sets 
of the Bethe parameters $\bar t^p$ for $p=1,\ldots,N-1$ are not empty. 
If $\bar t^{\ell'}=\varnothing$ for some $\ell'\not=\ell$ 
then for $\ell'>\ell$ the sum in \r{c7} over $j$ ends at $j=\ell'$ and for $\ell'<\ell$ 
the the sum over $i$ begins at  $\ell'+1$.

\begin{cor}\label{cor35}
There are two extreme cases of the recurrence relations \r{c7} 
with respect to first and last Bethe parameters when $\ell=1$ and $\ell=N-1$ 
\begin{equation}\label{rr1}
\begin{split}
\BB(\{z,\bar t^1\};\{\bar t^s\}_2^{N-1})&=\sum_{j=2}^N \frac{\Tt_{1,j}(z)}{\lambda_1(z)}
\sum_{{\rm part}}\BB(\bar t^1;\{\bar t^s_{\st}\}_2^{j-1};\{\bar t^s\}_j^{N-1})\times\\
&\quad \times \frac{1}{f(\bar t^2,z)} \prod_{p=2}^{j-1}
\frac{\gle(\bar t^p_{\sth},\bar t^{p-1}_{\sth})f(\bar t^p_{\st},\bar t^p_{\sth})}
{ f(\bar t^{p+1},\bar t^{p}_{\sth})}
\end{split}
\end{equation}
and 
\begin{equation}\label{rr2}
\begin{split}
\BB(\{\bar t^s\}_1^{N-2};\{z,\bar t^{N-1}\})&=\sum_{j=1}^{N-1} \frac{\Tt_{j,N}(z)}{\lambda_{N-1}(z)}
\sum_{{\rm part}}\BB(\{\bar t^s\}_1^{j-1};\{\bar t^s_{\st}\}_j^{N-2};\bar t^{N-1})\times\\
&\quad \times \frac{1}{f(z,\bar t^{N-2})}
\prod_{p=j}^{N-2}\frac{\beta_p(\bar t^p_{\so})
\gri(\bar t^{p+1}_{\so},\bar t^{p}_{\so})f(\bar t^p_{\so},\bar t^p_{\st})}
{f(\bar t^{p}_{\so},\bar t^{p-1})}\,.
\end{split}
\end{equation}
\end{cor}

Recurrence relations \r{rr1} and \r{rr2} were obtained in \cite{HLPRS18} in the framework 
of the hierarchical Bethe ansatz techniques and for different normalization
of the off-shell Bethe vectors described in appendix~\ref{ApC}. 
One can verify that these recurrence relations are compatible with the 
hierarchical relations for the Bethe vectors described in \cite{OPS}.

\section{Zero modes method}
\label{zmmeth}
 
 In this paper we present trigonometric version of the zero modes method developed for 
 rational case in \cite{HLPRS-RA}. This method uses the fact that all entries of the 
 monodromy matrix can be obtained from one selected entry using commutation relations with 
 zero modes similar to \r{zmc2} and \r{zmc3}. 
 Using the action of zero modes and any particular element of monodromy matrix onto 
 Bethe vector one can obtain an action of the rest entries by induction. 
 
The induction step is based on relations between  the action of monodromy matrix entry 
$\Tt_{i,j}(z)$
onto off-shell Bethe vector and  
the action of the entries $\Tt_{i+1,j}(z)$, $\Tt_{i,j-1}(z)$ 
and zero modes $\LL^-_{i+1,i}[0]$, $\LL^-_{j,j-1}[0]$ (see formulas \r{zmc2} and  \r{zmc3} below). 
These relations allow to obtain the action 
of the entries $\Tt_{i+1,j}(z)$, $\Tt_{i,j-1}(z)$ assuming the action of $\Tt_{i,j}(z)$. 
Since the action of the matrix entry $\Tt_{1,N}(z)$ can be calculated using 
explicit presentation of the off-shell Bethe vector $\BB(\bar t)$ \r{BV} 
one can prove by induction the action formulas of all entries  $\Tt_{i,j}(z)$ for all $1\leq i,j\leq N$. 
This is done in appendix~\ref{ApA}. The action of the zero mode 
operators $\LL^-_{\ell+1,\ell}[0]$ onto off-shell Bethe vectors are given by the 
proposition~\ref{zmacBV}  and the action of the entry $\Tt_{1,N}(z)$ is calculated 
in proposition~\ref{hTac}. 

To develop zero modes method one needs to use the commutation relations 
of the zero modes $\LL^-_{l+1,l}[0]$ with monodromy matrix entry 
$\Tt_{i,j}(z)\equiv \LL^+_{i,j}(z)$ for $l=i$ and $l=j-1$. Taking $v=0$ in 
\r{TM-1} for $\mu=+$, $\nu=-$ and $\{i,j,k,l\}\to\{i,j,i+1,i\}$ one obtains 
\begin{equation}\label{zmc2}
\begin{split}
& \Tt_{i,j}(z)\ \LL^-_{i+1,i}[0]- q^{\delta_{i,j}}\ \LL^-_{i+1,i}[0]\ \Tt_{i,j}(z)=\\
&\quad= (q-q^{-1})\Big(\delta_{i,j-1}\ \LL^-_{j,j}[0]\ \Tt_{i,i}(z)- 
 \Tt_{i+1,j}(z)\ \LL^-_{i,i}[0]  \Big).
\end{split}
\end{equation}
Analogously, taking $u=0$ in the commutation relation \r{TM-1} for $\mu=-$, $\nu=+$ and 
$\{i,j,k,l\}\to\{i,j,j,j-1\}$
one gets 
\begin{equation}\label{zmc3}
\begin{split}
&q^{-\delta_{i,j}}\ \LL^-_{j,j-1}[0]\ \Tt_{i,j}(z)\ -  \Tt_{i,j}(z)\ \LL^-_{j,j-1}[0] =\\
&\quad= (q-q^{-1})\Big(\delta_{i,j-1}\  \LL^-_{i,i}[0]\ \Tt_{j,j}(z)- 
\Tt_{i,j-1}(z)\ \LL^-_{j,j}[0]  \Big).
\end{split}
\end{equation}
Also we will use commutation relation with zero modes of diagonal entries of monodromy matrices. Taking $v=0$ in 
\r{TM-1} for $\mu=+$, $\nu=-$  and $\{i,j,k,l\}\to\{i,j,l,l\}$ one obtains
\begin{equation}\label{zmcd}
\begin{split}
 q^{\delta_{il}}  \Tt_{i,j}(z)\ \LL^-_{l,l}[0] = 
  q^{\delta_{lj}} \LL^-_{l,l}[0]\  \Tt_{i,j}(z) .
\end{split}
\end{equation}

According to the Gauss decomposition \r{Gauss1} and restrictions \r{restr}
\begin{equation}\label{zm4}
\LL^-_{i,i}[0]=k_i^{-1},\qquad \LL^-_{i+1,i}[0]=k_i^{-1}\ E_{i}\,,
\end{equation}
where $k_i$ and $E_i$ are zero modes operators of the Gauss coordinates 
$k_i^+(u)$ and $\EE^-_{i,i+1}(u)$ respectively: $k_i=k^+_i[0]=k^+_i(u)|_{u=\infty}$ and 
$E_i=\EE^-_{i,i+1}[0]=\EE^-_{i,i+1}(u)|_{u=0}$.

The action of the zero mode operators onto vacuum vector $\rvac$ is defined as follows
\begin{equation}\label{zm5}
k_i\rvac=\kappa_i^{-1}\rvac,\quad E_i\rvac =0,
\end{equation}
where $\kappa_i\in\CC$ are complex parameters .

The action of zero mode operators $\LL^-_{i+1,i}[0]$ onto off-shell Bethe vector
\r{BV} is given by the following 
\begin{prop}\label{zmacBV}
\begin{multline}\label{zm6}
\LL^-_{i+1,i}[0]\ \BB(\bar t)=(q-q^{-1})\times \\ \times\sum_{\ell=1}^{r_i}
\Big( \kappa_i\ q^{r_{i}-r_{i-1}-1}\ \beta_{i}(t^i_\ell)\ 
\frac{f(t^i_\ell,\bar t^{i}_\ell)}{f(t^i_\ell,\bar t^{i-1})}- \kappa_{i+1}\ q^{r_{i+1}-r_{i}+1}\ 
\frac{f(\bar t^i_\ell,t^{i}_\ell)}{f(\bar t^{i+1},\bar t^{i}_\ell)}\Big)\ \BB(\bar t\setminus\{t^i_\ell\}).
\end{multline}
\end{prop}

\proof To prove this proposition one can use the presentation of the off-shell
pre-Bethe vector \cite{FKPR08}
\begin{equation}\label{zm7}
\Pfp\Big(\FFF(\bar t)\Big)=\FFF(\bar t)+\sum_{i=1}^{N-1}\sum_{\ell=1}^{r_i}
\frac{f(\bar t^i_\ell,t^i_\ell)}{f(\bar t^{i+1},t^i_\ell)}\ \FF^-_{i+1,i}(t^i_\ell)\ \FFF(\bar t\setminus\{t^i_\ell\})
+\cdots
\end{equation}
where $\cdots$ stands for the terms which are annihilated by projection $\Pfp$ after 
adjoint action 
of zero mode operators $E_i$, $i=1,\ldots,N-1$. The commutation relations 
between Gauss coordinates for $\mu,\nu=\pm$
\begin{equation}\label{zm8}
[\EE^\mu_{i,i+1}(u),\FF^\nu_{j+1,j}(v)]=\delta_{i,j}\gle(v,u)\Big(k^\nu_{i+1}(v)k^\nu_{i}(v)^{-1}-
k^\mu_{i+1}(u)k^\mu_{i}(u)^{-1}\Big)
\end{equation}
imply the commutation relations
\begin{equation}\label{zm9}
[E_i,\FF^-_{i+1,i}(v)]=(q-q^{-1})\Big(k^-_{i+1}(v)k^-_{i}(v)^{-1}-k_{i+1}^{-1} k_i\Big)
\end{equation}
and 
\begin{equation}\label{zm10}
[E_i,F_{i}(v)]=(q-q^{-1})\Big(k^+_{i+1}(v)k^+_{i}(v)^{-1}-k^-_{i+1}(v)k^-_{i}(v)^{-1}\Big).
\end{equation}
Formulas \r{zmcd} yield  
\begin{equation}\label{zm11}
k_i\ F_i(v)\ k_i^{-1}= q^{-1}\ F_i(v),\quad k_{i+1}\ F_i(v)\ k_{i+1}^{-1}= q\ F_i(v)
\end{equation}
which imply 
\begin{equation}\label{zm12}
k^{-1}_{i}\ \FFF(\bar t)\ k_{i}
=q^{r_{i}-r_{i-1}}\ \FFF(\bar t)\,.
\end{equation}
Equality \r{zm10}  together with \r{kEFA}  yields 
\begin{equation}\label{zm13}
[E_i,\FFF(\bar t)]=(q-q^{-1})\sum_{\ell=1}^{r_i} 
\frac{f(t^i_\ell,\bar t^i_\ell)}{f(t^i_\ell,\bar t^{i-1})}\ \FFF(\bar t\setminus\{t^i_\ell\})\
k^+_{i+1}(t^i_\ell)k^+_{i}(t^i_\ell)^{-1}
\end{equation}
where we skip the terms proportional to $k^-_{i+1}(t^i_\ell)k^-_{i}(t^i_\ell)^{-1}$ since 
they will vanish after restriction of the action $\LL^-_{i+1,i}[0]\ \Pfp\Big(\FFF(\bar t)\Big)$
onto subalgebra $U^+_F=U_F\cap U^+_q(\glaf)$. 
Taking into account \r{zm4}, \r{zm5}, \r{zm7}, \r{zm9} and \r{zm12}
we obtain the statement of the proposition.  \qed

\begin{remark}
Note that unlike the rational quantum integrable models the on-shell Bethe vectors 
are not the highest weight vectors of $U_q(\mathfrak{gl}_N)$ 
with respect to the action of $E_i$. 
\end{remark} 

In proving the action formulas of the 
monodromy matrix entries onto off-shell Bethe vectors we will use induction to obtain
the action of the entries $T_{i+1,j}(z)$  and $T_{i,j-1}(z)$  
from the induction assumption of the action  $T_{i,j}(z)$. 
To do this we will equate the terms of \r{zmc2} and \r{zmc3} acting onto $\BB(\bar t)$ 
at the different parameters $\kappa_i$
using their arbitrariness.  
In order to  formulate the base of the induction 
one can prove the following  
\begin{prop}\label{hTac}
The  monodromy matrix element $\Tt_{1,N}(z)$ is acting onto off-shell Bethe 
vector $\BB(\bar t)$ as follows
\begin{equation}\label{zzm7}
\Tt_{1,N}(z)\ \BB(\bar t)=\lambda_1(z)\ \BB(\bar w),
\end{equation}
where $\bar w$ is a collection of sets of the variables $\{\bar w^1,\ldots,\bar w^{N-1}\}$ 
such that $\bar w^i=\{z,\bar t^i\}$ for $i=1,\ldots,N-1$.
\end{prop}

\proof To prove this proposition we need an auxiliary 
\begin{lemma}\label{auxlem}
There are the equalities in $U_q(\glaf)$
\begin{equation}\label{CGCp}
\FF^+_{N,1}(z)=\Pfp\sk{F_{N-1}(z)F_{N-2}(z)\cdots F_2(z)F_1(z)}
\end{equation}
and 
\begin{equation}\label{CGCm}
\Pfp\sk{\Tt_{1,N}(z)\cdot \FF^-_{j,i}(t)}=\Pfp\sk{\FF^+_{N,1}(z)k^+_1(z)\cdot \FF^-_{j,i}(t)}=0,
\quad \forall i,j\,.
\end{equation}
\end{lemma}
\proof The statement of the lemma can be proved using $\RR\LL\LL$ relation \r{TM-1}  (see 
appendix~D in \cite{LP21a}  for details).\qed

To calculate the action \r{zzm7} we will use \r{CGCp} and \r{CGCm}
\begin{equation}\label{GLT2}
\begin{split}
&\Tt_{1,N}(z)\cdot \BB(\bar t)= \Tt_{1,N}(z)\cdot \Pfp\sk{\FFF(\bar t)}\rvac=\Pfp\sk{ \Tt_{1,N}(z)\cdot \FFF(\bar t)}\rvac=\\
&\quad=\left.\Pfp\sk{F_{N-1}(z_{N-1})\cdots F_1(z_1)k^+_{1}(z_{1}) \FFF(\bar t)}\rvac\right|_{z_i=z}
=\lambda_{1}(z)\ X(z,\bar t)\ \BB(\bar w)\,,
\end{split}
\end{equation}
where 
\begin{equation*}
X(z,\bar t)=f(\bar t^1,z_1) \left.
\prod_{i=1}^{N-1}  \frac{\dfun_f(\bar t^i)}{\dfun_f(\bar w^i)}
\prod_ {i=1}^{N-2} \frac{f(\bar w^{i+1},\bar w^i)}{f(\bar t^{i+1},\bar t^i)f(z_{i+1},\bar w^i)}
\right|_{z_i=z} =1
\end{equation*}
and for $i=1,\ldots,N-1$
\begin{equation}\label{d-fun}
\dfun_f(\bar t^i)=\prod_{1\leq\ell<\ell'\leq |\bar t^i|} f(t^i_{\ell'},t^i_\ell)\,.
\end{equation}

In order to perform this calculation using the commutation relations between total currents \r{kEFA}
we split parameters $z_i\not=z_j$ and denote $\bar w^i=\{z_i,\bar t^i\}$. 
In \r{GLT2} we also use  property of the projections that \cite{EKhP07}
\begin{equation*}
\Pfp\sk{\FF^+_{N,1}(z)\FFF(\bar t)}=\Pfp\sk{\Pfp\sk{F_{N-1}(z)\cdots F_1(z)}\FFF(\bar t)}=
\Pfp\sk{F_{N-1}(z)\cdots F_1(z)\FFF(\bar t)}.
\end{equation*}
\qed

\section{Dual Bethe vector and scalar product}\label{sec:BV}

Definition of dual Bethe vectors $\CC(\bu)$ for $U_q(\glaf)$ and scalar product between  
$\CC(\bu)$ and $\BB(\bar t)$ was given in \cite{HLPRS18}.  The recurrence 
relations for the highest coefficient (HC) of the scalar product  corresponding to 
the extreme recurrence relations \r{rr1} and \r{rr2} were found in this paper. 
Here we extend this result 
to more general recurrence relation \r{c7} given by the proposition~\ref{rr1var}. 

Dual Bethe vectors are the vectors of the dual representation space $\mathcal{H}^*$   of the algebra $U_q(\glaf)$ which 
possesses a  vector $\lvac$ with properties similar to \r{vac-def}
\begin{equation}\label{vac-d}
\lvac\LL_{i,j}(u)=0,\quad i<j,\qquad \lvac \LL_{i,i}(u)=\lambda_i(u)\lvac\,.
\end{equation}

In order to define dual Bethe vectors we consider an  involutive antimorphism 
\begin{equation}\label{ant1}
\Psi:U_q(\glaf)\to U_{q^{-1}}(\glaf),\qquad \Psi(A\cdot B)=\Psi(B)\cdot \Psi(A)
\end{equation} 
defined on the matrix entries of the $\LL$-operators as follows
\begin{equation}\label{ant2}
\Psi\Big(\LL_{i,j}(u)\Big)=\tLL_{j,i}(u^{-1})
\end{equation}
where $\tLL_{i,j}(u)\in U_{q^{-1}}(\glaf)$.

The action of antimorphism \r{ant1} can be extended to the action onto vectors 
from  
representation spaces $\mathcal{H}$ and $\mathcal{H}^*$ according to the rules
\begin{equation}\label{ant3}
\begin{split}
\Psi(\rvac)=\tilde{\lvac},\qquad \Psi(A\rvac)=\tilde{\lvac}\Psi(A),\\
\Psi(\lvac)=\tilde{\rvac},\qquad \Psi(\lvac A)=\Psi(A)\tilde{\rvac}.
\end{split}
\end{equation}
Here vectors $\tilde{\rvac}$ and $\tilde{\lvac}$ are defined by \r{vac-def} and 
\r{vac-d} for the algebra $U_{q^{-1}}(\glaf)$ and $A$ is any product of the entries 
$\LL_{i,j}(z)$ of the $\LL$-operator.
Note also that according to the equalities
\begin{equation*}
 \Psi( \LL_{i,i}(z) \rvac ) = \lambda_{i}(z) \tilde{\lvac} = \tilde{\lvac} \
 \tilde{\LL}_{i,i}(z^{-1})  
 = \tilde{\lambda}_{i}(z^{-1}) \tilde{\lvac}
\end{equation*}
one has 
\begin{equation}\label{lameq}
\tilde{\lambda}_{i}(z^{-1})={\lambda}_{i}(z)\,.
\end{equation}
Here $\tilde{\lambda}_{i}(z)$ are eigenvalues of the diagonal entries of the  $\LL$-operator for the algebra $U_{q^{-1}}(\glaf)$.

It was demonstrated in \cite{HLPRS18} that the dual Bethe vectors 
can be defined as follows 
\begin{equation}\label{dBV}
\Psi\Big(\BB_q(\bar t)\Big)=\CC_{q^{-1}}(\bar t^{-1}),\qquad 
\Psi\Big(\CC_q(\bar t)\Big)=\BB_{q^{-1}}(\bar t^{-1})\,,
\end{equation}
where notation $\bar t^{-1}$ means
\begin{equation*}
\bar t^{-1}=\Big\{\frac{1}{t^1_1},\ldots,\frac{1}{t^1_{r_1}}; 
\frac{1}{t^2_1},\ldots,\frac{1}{t^2_{r_2}}; \ldots; 
\frac{1}{t^{N-1}_1},\ldots,\frac{1}{t^{N-1}_{r_{N-1}}}\Big\}
\end{equation*}
and subscript in notation of the Bethe vectors $\BB_{q^{-1}}(\bt)$ and 
$\CC_{q^{-1}}(\bt)$ signifies that they are constructed for the algebra 
$U_{q^{-1}}(\glaf)$.

The scalar product  of a generic  Bethe vector $\BB(\bt)$ and 
a generic dual Bethe vector $\CC(\bs)$ 
\begin{equation}\label{SP0}
 S(\bs | \bt) = \CC(\bs) \BB(\bt)
\end{equation}
is not vanishing unless $|\bs^i|=|\bt^i|$ for all $i=1,\ldots,N-1$.
The scalar product $ S(\bs | \bt)$ can be written in the form \cite{HLPRS18}
\begin{equation}\label{SP00}
 S(\bs | \bt) =
 \sum_{{\rm part}}  W( \bs_{\so},\bs_{\st}|\bt_{\so},\bt_{\st})
 \prod_{i=1}^{N-1} \beta_{i}(\bs^{i}_{{\st}})  \beta_{i} (\bt^{i}_{{\so}}),
\end{equation}
where sum goes over partitions $\{\bs^i_{\so},\bs^i_{\st}\}\vdash\bs^i$ and 
$\{\bt^i_{\so},\bt^i_{\st}\}\vdash\bt^i$ such that $|\bs^i_{\so}|={|\bt^i_{\so}|}$
for $i=1,\ldots,N-1$. The rational function $W( \bs_{\so},\bs_{\st}|\bt_{\so},\bt_{\st})$
depends only on the $\RR$-matrix of the model and does not depend on the 
free functional parameters $\lambda_i(z)$. 
The function
$W( \bs_{\so},\bs_{\st}|\bt_{\so},\bt_{\st})$ has a presentation 
 \cite{HLPRS18}  
\begin{equation}\label{SC1}
W( \bs_{\so},\bs_{\st}|\bt_{\so},\bt_{\st}) =
{Z} (\bs_{\so} |\bt_{\so} ) \  \bar{Z}(\bs_{\st}|\bt_{\st})\
 \frac{\prod_{i=1}^{N-1} f(\bs^{i}_{\st}, \bs^{i}_{\so}) f(\bt^{i}_{\so}, \bt^{i}_{\st}) }
      {\prod_{i=1}^{N-2} f(\bs^{i+1}_{\st}, \bs^{i}_{\so})
 f(\bt^{i+1}_{\so}, \bt^{i}_{\st})} ,
\end{equation}
where 
\begin{equation}\label{HC1}
 W(\bs, \varnothing |\bt, \varnothing) = {Z}(\bs | \bt ), \quad
 W(\varnothing, \bs | \varnothing, \bt) = \bar{Z}(\bs | \bt )
\end{equation}
are  so called {\it highest 
coefficient} of the scalar product. One can check that $\bar{Z}(\bu|\bt)=Z(\bt|\bu)$
due to the symmetry $S(\bs | \bt)=\Psi(S(\bs | \bt))=S(\bt | \bs)$.

Recurrence relation \r{c7}
implies the recurrence relation for the dual Bethe vectors 
\begin{multline}\label{SP3}
\CC(\{\bu^s\}_1^{\ell-1},\{\bu^\ell_{\so}, \bu^\ell_{\st}\},\{\bu^s\}_{\ell+1}^{N-1})=
\sum_{i=1}^\ell\sum_{j=\ell+1}^{N}\sum_{\rm part}
\frac{\CC(\{\bar u^{s} \}_1^{i-1},\{\bu^{s}_{\st}\}_i^{j-1},\{\bu^{s}\}_{j}^{N-1}))}
{f(\bu^\ell_{\so},\bu^{\ell-1})f(\bu^{\ell+1},\bu^\ell_{\so})}
\frac{T_{j,i}(\bu^\ell_{\so})}{\lambda_\ell(\bu^\ell_{\so})} \times \\
\times \prod_{p=i}^{\ell-1} \frac{\beta_p(\bu^p_{\so})
\gle(\bu^{p+1}_{\so},\bu^{p}_{\so})f(\bu^p_{\so},\bu^p_{\st})}
{f(\bu^{p}_{\so},\bu^{p-1})}\ \prod_{p=\ell+1}^{j-1}
\frac{\gri(\bu^p_{\sth},\bu^{p-1}_{\sth})f(\bu^p_{\st},\bu^p_{\sth})}
{ f(\bu^{p+1},\bu^{p}_{\sth})}
\end{multline}
where we used \r{lameq}, 
\begin{equation*}
f_{q^{-1}}(x^{-1},y^{-1})=f_q(x,y),\qquad \gle_{q^{-1}}(x^{-1},y^{-1})=\gri_q(x,y)
\end{equation*}
{and sum over partitions is the same as in \eqref{c7}.}
Note that there are no summation over partition of the set $\bu^\ell$ in 
 \r{SP3}. 

Let us consider the scalar product \r{SP0}
  for some fixed partition  $\{\bu^\ell_{\so}, \bu^\ell_{\st}\}\vdash \bu^\ell$ 
  with cardinality $|\bu^\ell_{\so}|=1$. 
  Then, one can prove following 
 \begin{prop}\label{HCprop}
 Highest coefficients  $Z(\bu|\bt)$ and 
 $\bar{Z}(\bu|\bt)$ satisfy the recurrence relations 
\begin{multline}\label{eq:HCrec2}
 {Z}(\bu|\bt)  = 
 \sum_{\rm part}
\frac{ f(\bt^\ell,\bu^{\ell}_{\so})}
{f(\bu^\ell_{\so},\bu^{\ell-1})f(\bu^{\ell+1},\bu^\ell_{\so})} \ 
\frac{\gle(\bt^\ell_{\so}, \bw^{\ell-1}_{\so}) f(\bt^\ell_{\so},\bt^\ell_{\st})}
     { f(\bt^\ell_{\so},\bar w^{\ell-1})} \\
\times  \sum_{j=\ell+1}^{N}
\prod_{p=1}^{\ell-1} 
 \frac{\gle(\bar w^p_{\so}, \bar w^{p-1}_{\so}) f(\bw^p_{\so},\bw^p_{\st})}
      { f(\bar w^p_{\so},\bar w^{p-1})}
\prod_{p=\ell+1}^{j-1} 
\frac{\gri(\bu^p_{\sth},\bu^{p-1}_{\sth})f(\bu^p_{\st},\bu^p_{\sth})}
{ f(\bu^{p+1},\bu^{p}_{\sth})}
 \frac{\gle(\bt^p_{\so}, \bt^{p-1}_{\so}) f(\bt^p_{\so},\bt^p_{\st})}
      { f(\bt^p_{\so},\bt^{p-1})} \\
\times {Z}(\{\bu^s\}_1^{\ell-1},\{\bu^s_{\st}\}_\ell^{j-1},\{\bu^s\}_{j}^{N-1}| \{\bw^s_{\st}\}_1^{\ell-1}, \{\bt^s_{\st}\}_\ell^{j-1}, \{\bt^s\}_{j}^{N-1})
\qquad
\end{multline}
and 
\begin{multline}\label{eq:HCrec1}
\bar{Z}(\bu | \bt) = 
\sum_{\rm part}
\frac{ f(\bu^{\ell}_{\so},\bar t^{\ell})}
{f(\bu^\ell_{\so},\bu^{\ell-1})f(\bu^{\ell+1},\bu^\ell_{\so})}\ 
 \frac{\gri(\bar w^{\ell+1}_{\sth}, \bt^{\ell}_{\sth})   f(\bt^\ell_{\st},\bt^\ell_{\sth})   }
	  { f(\bar w^{\ell+1}, \bt^{\ell}_{\sth})} \\
\times \sum_{i=1}^\ell
\prod_{p=i}^{\ell-1}  \frac{\gle(\bu^{p+1}_{\so},\bu^{p}_{\so})f(\bu^p_{\so},\bu^p_{\st})}
     {f(\bu^{p}_{\so},\bu^{p-1})}\ 
 \frac{\gri(\bt^{p+1}_{\sth},\bt^{p}_{\sth})f(\bt^p_{\st},\bt^p_{\sth})   }
	  { f(\bt^{p+1},\bt^{p}_{\sth})}    
\prod_{p=\ell+1}^{N-1} 
 \frac{\gri(\bar w^{p+1}_{\sth}, \bar w^{p}_{\sth})f(\bar w^p_{\st},\bar  w^p_{\sth})   }
      {f(\bar w^{p+1}, \bar w^{p}_{\sth})}  
	\\
\times \bar{Z}(\{\bu^s\}_1^{i-1},\{\bu^s_{\st}\}_i^{\ell},\{\bu^s\}_{\ell+1}^{N-1}| \{\bt^s\}_1^{i-1}, \{\bt^s_{\st}\}_i^{\ell},\{\bw^s_{\st}\}_{\ell+1}^{N-1} ).
\qquad
\end{multline} 
In \r{eq:HCrec2}
partition is going over $ \{\bu^s_{\st}, \bu^s_{\sth} \} \vdash \bu^s$ for $\ell + 1 \le s \le j-1$,  
$ \{\bt^s_{\so}, \bt^s_{\st} \} \vdash \bt^s$ for $\ell \le s \le j-1$ and $\{\bu^{\ell}_{\so}, \bt^{s}\} 
=  \{\bw^s_{\so}, \bw^s_{\st} \} \vdash \bw^s$ for $1 \le s \le \ell-1$. 
In \r{eq:HCrec1} partition is going over $ \{\bu^s_{\so}, \bu^s_{\st} \} \vdash \bu^s$ for 
$i \le s \le \ell-1$,  
$ \{\bt^s_{\st}, \bt^s_{\sth} \}  \vdash \bt^s$ for $i \le s \le \ell$ and $\{\bu^{\ell}_{\so}, \bt^{s}\} = 
 \{\bw^s_{\st}, \bw^s_{\sth} \}  \vdash \bw^s $ for $\ell + 1 \le s \le N-1$. 
In both formulas \r{eq:HCrec2} and \r{eq:HCrec1}
there are no partition over set $\bu^\ell$.
Boundary sets are defined as follows: 
$\bu^{0} = \bt^{0} = \bw^{0}_{\st} = \bu^{N} = \bt^{N} = \bw^{N}_{\st} = \varnothing$ 
and $\bw^{0}_{\so} = \bw^{N}_{\sth} = \bu^{\ell}_{\so}$.
\end{prop}

{\it Proof} of this proposition consists of the three steps.
\begin{enumerate}
 \item 
First, one substitutes recurrence relation for the dual Bethe vector \r{SP3} in  the right hand side of 
the equality
 \begin{equation}\label{lreq}
  \CC(\bu) \ \BB(\bt) =  
  \CC(\{\bu^s\}_1^{\ell-1},\{\bu^\ell_{\so}, \bu^\ell_{\st}\},\{\bu^s\}_{\ell+1}^{N-1}) \ \BB(t)
 \end{equation}
 for some fixed partition $\{\bu^\ell_{\so},\bu^\ell_{\st}\}\vdash \bu^\ell$. 
 \item Then, one uses the action of the monodromy matrix entry  
 $T_{j,i}(\bu^\ell_{\so})$ onto Bethe vector $\BB(\bt)$ according to the   
 proposition~\ref{sinac}. 
 \item 
 Finally, the  statement of the proposition follows from 
 comparing the coefficients at the products 
 $\prod_{i=1}^{N-1}\beta_{i}(\bt^i)$ and $\prod_{i=1}^{N-1}\beta_{i}(\bu^i)$ in both sides of \r{lreq} in the presentation \r{SP00}. \qed
\end{enumerate}

The extreme case $\ell = 1$ in \r{eq:HCrec2} takes the  form 
\begin{multline}\label{HCrec2}
 {Z}(\bu|\bt)  = 
 \sum_{\rm part}\sum_{j=2}^{N}
\frac{\gle(\bt^1_{\so}, \bu^{1}_{\so})f(\bt^1_{\st},\bu^{1}_{\so})f(\bt^1_{\so},\bt^1_{\st})}
     {f(\bu^{2},\bu^{1}_{\so})}
\prod_{p=2}^{j-1}
\frac{\gri(\bu^p_{\sth},\bu^{p-1}_{\sth})f(\bu^p_{\st},\bu^p_{\sth})}
     {f(\bu^{p+1},\bu^{p}_{\sth})} 
 \frac{\gle(\bt^p_{\so}, \bt^{p-1}_{\so})f(\bt^p_{\so},\bt^p_{\st})}
      { f(\bt^p_{\so},\bt^{p-1})} \\
{Z}( \{\bu^s_{\st}\}_1^{j-1},\{\bu^s\}_{j}^{N-1}|  \{\bt^s_{\st}\}_1^{j-1}, \{\bt^s\}_{j}^{N-1}), 
\qquad\qquad\qquad 
\end{multline}
where partition is going over $\{\bu^s_{\so}, \bu^s_{\st} \}  \vdash \bu^s$ for $2 \le s \le j-1$,  
$ \{\bt^s_{\st}, \bt^s_{\sth} \}  \vdash \bt^s$ for $i \le 1 \le j-1$. 
One can check that this equality coincide identically with 
 formula (4.15) from \cite{HLPRS18}.

The extreme case $\ell = N - 1$ in \r{eq:HCrec1} takes the form 
\begin{multline}\label{HCrec1}
 \bar{Z}(\bu | \bt) = 
\sum_{i=1}^{N-1}
\sum_{\rm part}
\frac{{\gri}(\bu^{N-1}_{\so},\bt^{N-1}_{\sth})
f(\bu^{N-1}_{\so}, \bt^{N-1}_{\st})f(\bt^{N-1}_{\st},\bt^{N-1}_{\sth}) }
{f(\bu^{N-1}_{\so},\bu^{N-2})} 	  \\
\times\prod_{p=i}^{N-2}  
 \frac{{\gri}(\bt^{p+1}_{\sth},\bt^{p}_{\sth})f(\bt^p_{\st},\bt^p_{\sth})   }
	  {f(\bt^{p+1},\bt^{p}_{\sth})}  
 \frac{{\gle}(\bu^{p+1}_{\so},\bu^{p}_{\so})f(\bu^p_{\so},\bu^p_{\st})}
     {f(\bu^{p}_{\so},\bu^{p-1})}	  \\
 \times    \bar{Z}(\{\bu^s\}_1^{i-1},\{\bu^s_{\st}\}_i^{N-1}| \{\bt^s\}_1^{i-1}, \{\bt^s_{\st}\}_i^{N-1} ),
\qquad\qquad
\end{multline} 
where partition is going over $ \{\bu^s_{\so}, \bu^s_{\st} \} \vdash \bu^s$ for $i \le s \le N-2$,  
$ \{\bt^s_{\st}, \bt^s_{\sth} \} \vdash \bt^s$ for $i \le s \le N - 1$.  Again,  
equality \r{HCrec1} coincide identically with 
 formula (4.16) from \cite{HLPRS18} if one takes into account equality $\bar{Z}(\bu | \bt)=
 {Z}(\bt | \bu)$.

\section*{Conclusion}

This paper is a continuation of the research started in \cite{HLPRS-RA} 
to develop zero modes method for the quantum integrable models defined
by $U_q(\mathfrak{gl}_N)$-invariant $\RR$-matrices. The aim of this method 
is to find the action formulas of monodromy matrix entries onto off-shell Bethe vectors.
These action formulas can be further used to obtain various recurrence 
relations for the Bethe vectors itself and different physical quantities 
in this class of the quantum integrable models. In \cite{HLPRS-RA} the 
recurrence relations for the highest coefficients of the scalar products 
of the off-shell Bethe vectors were obtained for the rational integrable models 
associated to super-Yangian double $DY(\mathfrak{gl}(m|n)$. 
In this paper we extend these results to the 
quantum integrable models based on the quantum loop algebra 
$U_q(\glaf)$.

\section*{Acknowledgments}
The work of SP was supported in part by the RFBR Grant 19-01-00726-a.

\appendix

\section{Proof of the proposition~\ref{sinac}}\label{ApA}

The statement of this proposition is obviously equivalent to \r{zm7} from 
the proposition~\ref{hTac} for $i=1$ and $j=N$. Assume that it is valid also 
for arbitrary values $i$ and $j$ and prove using \r{zmc2} that it is valid as well for the action 
of $\Tt_{i+1,j}(z)$.  We will consider only the case of $j\geq i+1$ in details, since 
the case $j<i+1$ can be considered analogously.
\smallskip

Due to 
\begin{equation}\label{zzm8}
\LL^-_{i,i}[0]\BB(\bar t)=\kappa_i\ q^{r_{i}-r_{i-1}}\ \BB(\bar t)\,,
\end{equation}
 the action of \r{zmc2} onto off-shell Bethe vector 
$\BB(\bar t)$  transforms into 
\begin{equation}\label{zzm9}
\begin{split}
&\Big(\LL^-_{i+1,i}[0]\Tt_{i,j}(z)-\Tt_{i,j}(z)\LL^-_{i+1,i}[0]\Big) \BB(\bar t)=\\
&\quad =
(q-q^{-1})\Big(q^{r_{i}-r_{i-1}}\kappa_i\ \Tt_{i+1,j}(z)\ \BB(\bar t)-
\delta_{i+1,j}q^{r_{i+1}-r_{i}}\kappa_{i+1}\ \Tt_{i,i}(z)\ \BB(\bar t)\Big),
\end{split}
\end{equation} 
where we used commutativity $[\LL^-_{i+1,i+1}[0],\Tt_{i,i}(z)]=0$.
To obtain the action $\Tt_{i+1,j}(z)$ onto off-shell Bethe vector 
$\BB(\bar t)$ from \r{zzm9} we equate the terms 
proportional to $\kappa_i$ and check that the terms proportional to $\kappa_{i+1}$ 
are cancel each other in the left hand side of \r{zzm9}
for $j>i+1$ and yields the action of $\Tt_{i,i}(z)\ \BB(\bar t)$
for $j=i+1$.

For $j\geq i+1$ in the action \r{ac2} 
 $\bar w^s=\{z,\bar t^s\}=\bar w^s_{\st}$ and $\bar w^s_{\so}=\bar w^s_{\sth}=\varnothing$
for $s=i,\ldots,j-1$.
Using \r{zm6}  one gets
\begin{equation}\label{zzm10}
\begin{split}
\Tt_{i,j}(z)\ \LL^-_{i+1,i}[0]\ \BB(\bar t)\Big|_{\kappa_{i+1}=0}&=(q-q^{-1})\kappa_i\ 
q^{r_{i}-r_{i-1}-1}\lambda_1(z)\times\\
&\quad \times \sum_{\ell=1}^{r_i}\beta_i(t^i_\ell)
\frac{f(t^i_\ell,\bar t^i_\ell)}{f(t^i_\ell,\bar t^{i-1})}
\sum_{{\rm part}} \mathbb{B}(\bar w_{\st} )     \Act_{i,j}(\bar w_{\so};\bar w_{\st};\bar w_{\sth})\,,
   \end{split}
\end{equation}
where $\bar w^i_{\st}=\{z,\bar t^i_\ell\}$. Acting by the same operators in inverse order 
one gets
\begin{equation}\label{zzm11}
\begin{split}
 \LL^-_{i+1,i}[0]\ \Tt_{i,j}(z)\ \BB(\bar t)\Big|_{\kappa_{i+1}=0}&=(q-q^{-1})\kappa_i\ 
q^{r_{i}-r_{i-1}}\lambda_1(z)\times\\
&\times \sum_{{\rm part}} \mathbb{B}(\bar w_{\st} )   \beta_i(\bar w^i_{\so})
\frac{f(\bar w^i_{\so},\bar w^i_{\st})}{f(\bar w^i_{\so},\bar w^{i-1}_{\st})}         
  \Act_{i,j}(\bar w_{\so};\bar w_{\st};\bar w_{\sth})\,,
\end{split}
\end{equation}
where the set $\bar w^i=\{z,\bar t^i\}$ is divided into subsets $\{\bar w^i_{\so},\bar w^i_{\st}\}
\vdash \bar w^i$ such that $|\bar w^i_{\so}|=1$. In \r{zzm11} one should take into account 
that $|\bar w^{i-1}_{\st}|=r_{i-1}$ while $|\bar w^{i}_{\st}|=|\bar w^{i}|=r_{i}+1$.

The sum over $\ell$ in \r{zzm10} can be rewritten as sum over  partitions. 
One can replace the ratio 
\begin{equation*}
\frac{f(t^i_\ell,\bar t^i_\ell)}{f(t^i_\ell,\bar t^{i-1})}=
\frac{f(t^i_\ell,\bar w^i_{\st})}{f(t^i_\ell,\bar w^{i-1})}\,,
\end{equation*}
where $\bar w^i_{\st}=\{z,\bar t^i_\ell\}$ and add to the sum over $\ell$ the zero term 
proportional to 
\begin{equation*}
\frac{f(z,\bar t^i)}{f(z,\bar w^{i-1})}=0.
\end{equation*}
The sum over $\ell$ transforms into 
\begin{equation*}
\sum_{\ell=1}^{r_i}\frac{f(t^i_\ell,\bar t^i_\ell)}{f(t^i_\ell,\bar t^{i-1})}\ (\ \cdot\ )=
\sum_{{\rm part}}
\frac{f(\bar w^i_{\so},\bar w^i_{\st})}{f(\bar w^i_{\so},\bar w^{i-1}_{\st})}
\frac{1}{f(\bar w^i_{\so},\bar w^{i-1}_{\so})}\ (\ \cdot\ )\,,
\end{equation*}
where sum runs over partitions $\{\bar w^i_{\so},\bar w^i_{\st}\}
\vdash \bar w^i$ such that $|\bar w^i_{\so}|=1$.
Subtracting \r{zzm10} from \r{zzm11} and using \r{h-p1} one gets the action \r{ac2} 
with index $i$ replaced by $i+1$. 

Let us check that for $j>i+1$ the terms proportional to $\kappa_{i+1}$ in \r{zzm9} vanish.
The terms which are proportional to $\kappa_{i+1}$ in the action of 
$T_{i,j}(z)\ \LL^-_{i+1,i}[0]\ \BB(\bar t)$ is 
\begin{equation}\label{zzm12}
\begin{split}
\Tt_{i,j}(z)\ \LL^-_{i+1,i}[0]\ &\BB(\bar t)\Big|_{\kappa_{i}=0}=-(q-q^{-1})\kappa_{i+1}\ 
q^{r_{i+1}-r_{i}}\lambda_1(z)  \times\\
&\quad\times
\sum_{{\rm part}} \mathbb{B}(\bar w_{\st} )  \Act_{i,j}(\bar w_{\so};\bar w_{\st};\bar w_{\sth})
 \frac{f(\bar w^i_{\st},\bar w^i_{\sth})}{f(\bar w^{i+1}_{\st},\bar w^i_{\sth})}
 \frac{q}{f(\bar w^{i+1}_{\sth},\bar w^i_{\sth})}\,,
\end{split}
\end{equation}
where we again present the sum over $\ell$ after the action of zero mode operator 
$\LL^-_{i+1,i}[0]$ as sum over partitions of the set 
$\{\bar w^i_{\st},\bar w^i_{\sth}\}\vdash \bar w^i=\{z,\bar t^i\}$ with $|\bar w^i_{\sth}|=1$.
The action in inverse order can be written in the same form  as \r{zzm12} since 
$|\bar w^{i}_{\st}|=r_{i}$  and $|\bar w^{i+1}_{\st}|=r_{i+1}+1$ and coefficient at 
$\kappa_{i+1}$ in the left hand side of \r{zzm9} vanishes. 

In the case $j=i+1$ the action $\Tt_{i,i+1}(z)\ \LL^-_{i+1,i}[0]\ \BB(\bar t)$ is given 
by equality \r{zzm12}, where the set $\bar w^{i+1}$ is divided into subsets 
$\bar w^{i+1}_{\st}$ and $\bar w^{i+1}_{\sth}$ with cardinality $|\bar w^{i+1}_{\st}|=r_{i+1}$.
The action in inverse order $\LL^-_{i+1,i}[0]\ \Tt_{i,i+1}(z)\ \BB(\bar t)$
is 
\begin{equation}\label{zzm13}
\begin{split}
\LL^-_{i+1,i}[0]\ \Tt_{i,i+1}(z)\ \BB(\bar t)\Big|_{\kappa_{i}=0}&=-(q-q^{-1})\kappa_{i+1}\ 
q^{r_{i+1}-r_{i}}\lambda_1(z)  \times\\
&\quad\times
\sum_{{\rm part}} \mathbb{B}(\bar w_{\st} )  \Act_{i,i+1}(\bar w_{\so};\bar w_{\st};\bar w_{\sth})
 \frac{f(\bar w^i_{\st},\bar w^i_{\sth})}{f(\bar w^{i+1}_{\st},\bar w^i_{\sth})}\,.
\end{split} 
\end{equation}
Subtracting \r{zzm12} at $j=i+1$ from \r{zzm13} and using \r{h-p2}
one gets from \r{zzm9} the action of the diagonal monodromy entry $T_{i,i}(z)$ onto 
off-shell Bethe vector $\BB(\bar t)$.  \qed

\section{Proof of the proposition~\ref{rr1var}}\label{ApC}

In this appendix  we prove recurrence relations for the off-shell Bethe vectors 
$\BB(\bar t)$ formulated in the proposition~\ref{rr1var}. Particular cases of these 
relations mentioned in the corollary~\ref{cor35} were  
found in \cite{HLPRS18} in the framework of hierarchical Bethe ansatz. 
Note that   in \cite{HLPRS18} normalization of the off-shell Bethe vectors was 
\begin{equation}\label{HLPRS-BV}
\prod_{i=1}^{N-1}\beta_i(\bar t^i)^{-1}\ \BB(\bar t)\,.
\end{equation}

 To prove proposition~\ref{rr1var} one can use the action formulas \r{ac2} and verify that 
 \r{c7}  becomes identities. Since the monodromy matrix entries 
 $T_{i,j}(z)$ in \r{c7} are upper-triangular $i<j$ we  substitute in the 
right hand side of this equality the action 
given by \r{acsim}:
\begin{equation}\label{rr3}
\begin{split}
\sum_{{\rm part}}\sum_{i=1}^{\ell}\sum_{j=\ell+1}^N
&\frac{\lambda_1(z)}{\lambda_\ell(z)}\ 
\frac{\BB(\{\bar w^s_{\st}\}_1^{i-1}, \{z,\bar t^s_{\st}\}_i^{\ell-1},
\{z,\bar t^\ell\};\{z,\bar t^s_{\st}\}_{\ell+1}^{j-1};\{\bar w^s_{\st}\}_j^{N-1})}
{f(\bar t^{\ell+1},z)f(z,\bar t^{\ell-1})}\times\\
&\quad\times 
\prod_{p=1}^{i-1} \frac{\beta_p(\bar w^p_{\so}) 
f(\bar w^p_{\so},\bar w^p_{\st})}
   {\hle(\bar w^p_{\so}, \bar w^{p-1}_{\so}) f(\bar w^p_{\so},\bar w^{p-1}_{\st})}
\prod_{p=i}^{\ell-1} \frac{\beta_p(\bar t^p_{\so})
\gri(\bar t^{p+1}_{\so},\bar t^{p}_{\so})f(\bar t^p_{\so},\bar t^p_{\st})}
{f(\bar t^{p}_{\so},\bar t^{p-1})}\times\\
 &\quad\times  \prod_{p=\ell+1}^{j-1}
\frac{\gle(\bar t^p_{\sth},\bar t^{p-1}_{\sth})f(\bar t^p_{\st},\bar t^p_{\sth})}
{ f(\bar t^{p+1},\bar t^{p}_{\sth})}
\prod_{p=j}^{N-1}      \frac{  f(\bar w^p_{\st},\bar w^p_{\sth})   }
	{\hri(\bar w^{p+1}_{\sth}, \bar w^{p}_{\sth})f(\bar w^{p+1}_{\st}, \bar w^{p}_{\sth})}\,,
\end{split}
\end{equation}
where sum over partitions of the sets 
$\{\bar t^s\}_i^{\ell-1}$,
$\{\bar t^s\}_{\ell+1}^{j-1}$ are described in propositions~\ref{rr1var} 
and of the sets $\{\bar w^s\}_1^{i-1}$, $\{\bar w^s\}_j^{N-1}$ in 
proposition~\ref{sinac} respectively. Recall that according to these rules 
$\bar w^0_{\so}=\bar t^\ell_{\so}=
\bar t^{\ell}_{\sth}=\bar w^N_{\sth}=\{z\}$. Our goal is to rewrite the sum over 
partitions of the sets $\{\bar t^s\}_{i}^{\ell-1}$ and 
$\{\bar t^s\}_{\ell+1}^{j-1}$ in \r{rr3} as sum over partitions 
of the sets $\{\bar w^s\}_i^{\ell-1}=\{z,\bar t^s\}_i^{\ell-1}$ and 
 $\{\bar w^s\}_{\ell+1}^{j-1}=\{z,\bar t^s\}_{\ell+1}^{j-1}$.

 To do this we transform  
second and third lines of \r{rr3} as follows 
\begin{equation}\label{rr4}
\begin{split}
&\quad\times 
\prod_{p=1}^{i-1} \frac{\beta_p(\bar w^p_{\so}) 
f(\bar w^p_{\so},\bar w^p_{\st})}
   {\hle(\bar w^p_{\so}, \bar w^{p-1}_{\so}) f(\bar w^p_{\so},\bar w^{p-1}_{\st})}
\prod_{p=i}^{\ell-1} \frac{\beta_p(\bar t^p_{\so})
\gri(\bar t^{p+1}_{\so},\bar t^{p}_{\so})f(\bar t^p_{\so},\{z,\bar t^p_{\st}\})}
{f(\bar t^{p}_{\so},\{z,\bar t^{p-1}\})}\times\\
 &\quad\times  \prod_{p=\ell+1}^{j-1}
\frac{\gle(\bar t^p_{\sth},\bar t^{p-1}_{\sth})f(\{z,\bar t^p_{\st}\},\bar t^p_{\sth})}
{ f(\{z,\bar t^{p+1}\},\bar t^{p}_{\sth})}
\prod_{p=j}^{N-1}      \frac{\gri(\bar w^{p+1}_{\sth}, \bar w^{p}_{\sth})
  f(\bar w^p_{\st},\bar w^p_{\sth})   }
	{f(\bar w^{p+1}, \bar w^{p}_{\sth})}\,.
	\end{split}
\end{equation}

Let us at the moment split the values of the boundary sets 
$\bar t^\ell_{\so}=\bar t^\ell_{\sth}=\{z'\}$ and $\bar w^0_{\so}=
\bar w^N_{\sth}=\{z\}$ with $z\not=z'$. 
Then the sum over partitions in \r{rr3} can be written as sum over 
partitions 
$\{\bar w^s_{\so},\bar w^s_{\st}\}_1^{\ell-1}
\vdash\{\bar w^s\}_1^{\ell-1}=\{z,\bar t^s\}_1^{\ell-1}$
and 
$\{\bar w^s_{\st},\bar w^s_{\sth}\}_{\ell+1}^{N-1}
\vdash\{\bar w^s\}_{\ell+1}^{N-1}=\{z,\bar t^s\}_{\ell+1}^{N-1}$
\begin{equation}\label{rr5}
\begin{split}
\sum_{{\rm part}}
&\frac{\BB(\{\bar w^s_{\st}\}_1^{\ell-1},\{z',\bar t^\ell\},\{\bar w^s_{\st}\}_{\ell+1}^{N-1})}
{f(\bar t^{\ell+1},z')f(z',\bar t^{\ell-1})}\times\\
&\quad\times \frac{\lambda_1(z)}{\lambda_\ell(z)}
\sum_{i=1}^{\ell}
\prod_{p=1}^{i-1}  \gle(\bar w^p_{\so}, \bar w^{p-1}_{\so}) 
\prod_{p=i}^{\ell-1} \gri(\bar w^{p+1}_{\so},\bar w^{p}_{\so})
\prod_{p=1}^{\ell-1}
\frac{\beta_p(\bar w^p_{\so})  f(\bar w^p_{\so},\bar w^p_{\st})}
   { f(\bar w^p_{\so},\bar w^{p-1})}
\times\\
 &\quad\times \sum_{j=\ell+1}^N \prod_{p=\ell+1}^{j-1}\gle(\bar w^p_{\sth},\bar w^{p-1}_{\sth})
\prod_{p=j}^{N-1}   \gri(\bar w^{p+1}_{\sth}, \bar w^{p}_{\sth})  
\prod_{p=\ell+1}^{N-1}
\frac{f(\bar w^p_{\st},\bar w^p_{\sth})}
{ f(\bar w^{p+1},\bar w^{p}_{\sth})}\,,
\end{split}
\end{equation}
where the boundary sets $\bar w^\ell_{\so}=\bar t^\ell_{\so}=\{z'\}$ and 
$\bar w^\ell_{\sth}=\bar t^\ell_{\sth}=\{z'\}$ are fixed. 
The terms in the sum over partitions vanishes
when either 
$\bar w^s_{\so}=\{z\}$ for $s=i,\ldots,\ell-1$ or
$\bar w^s_{\sth}=\{z\}$ for $s=\ell+1,\ldots,j-1$ 
 because either $f(z,\bar w^{s-1})^{-1}=0$ 
or $f(\bar w^{s+1},z)^{-1}=0$. 

Using a trivial relation 
between rational functions 
\begin{equation*}
\gri(x,y)=\gle(x,y)\ \frac{y}{x}
\end{equation*}
one can calculate the sums over $i$ and $j$ in  \r{rr5} to obtain  
\begin{equation}\label{rr6}
\begin{split}
&\sum_{{\rm part}}
\frac{\BB(\{\bar w^s_{\st}\}_1^{\ell-1},\{z',\bar t^\ell\},\{\bar w^s_{\st}\}_{\ell+1}^{N-1})}
{f(\bar t^{\ell+1},z')f(z',\bar t^{\ell-1})} 
\frac{\gle(z',\bar w^{\ell-1}_{\so})}{\gle(z',z)}
\frac{\gle(\bar w^{\ell+1}_{\sth},z')}{\gle(z,z')}\times\\
&\quad\times
\frac{\lambda_1(z)}{\lambda_\ell(z)}
\prod_{p=1}^{\ell-1}
\frac{\beta_p(\bar w^p_{\so})f(\bar w^p_{\so},\bar w^p_{\st})}
{f(\bar w^{p}_{\so},\bar w^{p-1}_{\st})
\hle(\bar w^{p}_{\so},\bar w^{p-1}_{\so})}
 \prod_{p=\ell+1}^{N-1}
\frac{f(\bar w^p_{\st},\bar w^p_{\sth})}
{f(\bar w^{p+1}_{\st},\bar w^{p}_{\sth})\hle(\bar w^{p+1}_{\sth},\bar w^p_{\sth})}\,.
\end{split}
\end{equation}
When $z'\to z$ in this sum over partitions only the term survives such that 
$\bar w^s_{\so}=\{z\}$ for all $s=1,\ldots,\ell-1$
and 
$\bar w^s_{\sth}=\{z\}$ for all $s=\ell+2,\ldots,N-1$. Taking into account 
that according to the proposition~\ref{sinac} 
$\bar w^0_{\st}=\bar w^N_{\st}=\varnothing$ and the fact that 
\begin{equation*}
\prod_{p=1}^{\ell-1}\beta_p(z)=\frac{\lambda_\ell(z)}{\lambda_1(z)}
\end{equation*}
one concludes that \r{rr6} is equal to 
$\BB(\{\bar t^s\}_{1}^{\ell-1},\{z,\bar t^\ell\},\{\bar t^s\}_{\ell+1}^{N-1})$ 
so \r{c7} becomes identity.  \qed


\begin{thebibliography}{99}

\bibitem{FadLH96} L. D. Faddeev, \textit{How Algebraic Bethe Ansatz works
for integrable model}, in: Les Houches Lectures \textsl{Quantum Symmetries}, eds A. Connes
et al, North Holland, (1998) 149, \texttt{arXiv:hep-th/9605187}.

\bibitem{Dr88} V.~G.~Drinfeld. \textsl{Quantum groups},
J. Soviet Math., {\bf 41}:2 (1988) 898--915.

\bibitem{RS1990}
N.Yu. Reshetikhin, M.A. Semenov-Tian-Shansky. {\sl Central extension of quantum groups}. 
Lett. Math. Phys. {\bf 19} (1990) 133--142. 

\bibitem{HLPRS-RA} A.~Hutsalyuk, A.~Liashyk,
S.~Z.~Pakuliak, E.~Ragoucy, N.~A.~Slavnov. {\sl Actions of the monodromy matrix elements 
onto $\mathfrak{gl}(m|n)$-invariant Bethe vectors}. J. Stat. Mech. (2020) 093104. 

\bibitem{HLPRS18} A.~Hutsalyuk, A.~Liashyk,
S.~Z.~Pakuliak, E.~Ragoucy, N.~A.~Slavnov. {\sl Scalar products and norm
of Bethe vectors for integrable models based on $U_q(\widehat{\mathfrak{gl}}_m)$}. 
SciPost (2018) {\bf 4} 006.

\bibitem{OPS} A. Os'kin, S. Pakuliak, A. Silantyev, {\sl On the universal weight function
for the quantum affine algebra $U_q(\mathfrak{gl}(N))$,}
Algebra and Analysis  {\bf 21} n.4 (2009)   196--240.


\bibitem{HLPRS17} A.~Hutsalyuk, A.~Liashyk,
S.~Z.~Pakuliak, E.~Ragoucy, N.~A.~Slavnov. {\sl Current presentation for the super-Yangian
double $DY(\mathfrak{gl}(m|n))$ and Bethe vectors}. Russ. Math. Surv.  
 (2017) {\bf 72} 33, doi:10.1070/RM9754.


\bibitem{LP21}
A. Liashyk, S. Z. Pakuliak.
{\sl Algebraic Bethe ansatz for $\mathfrak{o}_{2n+1}$-invariant integrable models.}
Theor. and Math. Phys. {\bf 206}(1) (2021) 19--39.



\bibitem{DF93} J.~Ding, I.~Frenkel. \textsl{Isomorphism of two realizations 
of quantum affine algebra
$U_q(\mathfrak{gl}(n))$}, Comm. Math. Phys. {\bf 156} (1993), 277--300,
{\tt DOI: 10.1007/BF02098484}


\bibitem{D88} V.~G.~Drinfeld. \textsl{A new realization of 
Yangians and of quantum affine algebras}, Soviet Math. Dokl. {\bf 36}
(1988) 212--216.

\bibitem{KhT93} Khoroshkin, S., Tolstoy, V. {\sl On Drinfeld realization of quantum affine algebras}.
Journal of Geometry and Physics {\bf 11} (1993), 101--108,
{\tt DOI: 10.1016/0393-0440(93)90070-U}


\bibitem{EKhP07} B.~Enriquez, S.~Khoroshkin, S.~Pakuliak, \textsl{Weight
functions and Drinfeld currents,}
{Comm. Math. Phys.} {\bf 276} (2007) 691--725.



\bibitem{LP21a} A.~Liashyk, S.~Z.~Pakuliak.
{\sl On the R-matrix realization of quantum loop algebras.}
\texttt{arXiv:2106.10666}

\bibitem{KhP-Kyoto} S. Khoroshkin, S. Pakuliak, {\sl A computation 
of an universal weight function for
the quantum affine algebra $U_q(\mathfrak{gl}(N))$},  {J. of Mathematics of Kyoto University},
{\bf 48} n.2 (2008) 277--321.





\bibitem{FKPR08} 
L. Frappat, S. Khoroshkin, S. Pakuliak, E. Ragoucy, 
\textsl{Bethe Ansatz for the Universal Weight Function},
Ann. H. Poincarre {\bf 10} (2009) 513, \texttt{arXiv:0810.3135}.






\end{thebibliography}
\end{document}